\documentclass[11pt]{article}

\usepackage{graphicx}
\usepackage{multirow}
\usepackage{amsmath,amssymb,amsfonts}
\usepackage{amsthm}
\usepackage{mathrsfs}
\usepackage[title]{appendix}
\usepackage{xcolor}
\usepackage{textcomp}
\usepackage{manyfoot}
\usepackage{booktabs}
\usepackage{algorithm}
\usepackage{algorithmicx}
\usepackage{algpseudocode}
\usepackage{listings}
\usepackage[utf8]{inputenc}
\usepackage{hyperref}
\begin{document}

\title{\textbf{Black Hole Evaporation Driven by Non-Thermal Squeezing Through SNS and CSNS Dynamics}}

\author{
Dhwani Gangal$^{1}$,
K.\,K. Venkataratnam$^{1}$\thanks{Corresponding author: \texttt{kvkamma.phy@mnit.ac.in}}
}

\date{
$^{1}$Department of Physics, Malaviya National Institute of Technology Jaipur,\\
J.\,L.\,N. Marg, Jaipur--302017, India
}

\maketitle

\begin{abstract}
In this work, we present a comprehensive semiclassical analysis of black hole radiation in a spatially flat FRW Universe for two fundamental nonclassical states: the Squeezed Number State (SNS) and the Coherent Squeezed Number State (CSNS). Unlike thermally modified earlier studies, SNS and CSNS constitute fully non-thermal, number-state-dependent quantum configurations. By embedding these states within the framework of semiclassical theory of gravity, we derive state-resolved expressions for the Hawking temperature, entropy variation, and corresponding mass loss of an evaporating black hole. The influence of the squeezing parameter $\rho$ and number state parameter $n$ on Hawking emission is examined through a series of analytical results supported by twelve detailed plots. The analysis reveals that the Hawking temperature exhibits monotonic growth with increasing $\rho$ and $n$, thereby elevating the effective temperature experienced at the black hole horizon. The entropy variations $\Delta \mathbb{S}_{\mathrm{SNS}}$ and $\Delta \mathbb{S}_{\mathrm{CSNS}}$ show strong nonlinear enhancement, especially at moderate and large squeezing values. Overall, the study extends earlier thermal squeezed-state approaches to a fully number-state-resolved framework and highlights the sensitivity of Hawking emission to nonclassical quantum configurations. These findings contribute a new perspective on gravitational particle creation in cosmological settings.
\end{abstract}

\vspace{0.5em}

\section{Introduction}
Modern cosmology seeks to understand evolution, origin, quantum structure etc of Universe within a framework that consistently connects classical gravity with quantum field theory. This theory supported by CMB observations, suggests that the early Universe was well approximated by a spatially flat FRW geometry \cite{kubik_origin_2022, moore_big_2014}. Although the Big Bang paradigm successfully describes large scale expansion, it does not resolve several foundational issues such as the horizon, flatness, monopole, and initial singularity problems. Inflationary theory first proposed by Guth \cite{guth1981cosmological}addresses these shortcomings by invoking a brief period of rapid accelerated expansion driven by a massive homogeneous scalar field known as the inflaton \cite{pedrosa_gaussian_2009, gangal2025Mandel}.

During the inflation period, energy density of the Universe is dominated by slowly varying inflaton potential, resulting in quasi exponential expansion \cite{geralico_novel_2004}. As inflation ends, the inflaton undergoes damped oscillation around the minimum of its potential, leading to the production of particles, reheating, and the subsequent restoration of thermal equilibrium \cite{green_cosmological_2022, albrecht_inflation_1994, yadav2024reheating, kofman_reheating_1994, allahverdi_reheating_2010}. This transition is central to understanding standard matter production, and it can be quantitatively characterized using reheating parameters \cite{yadav2024mutated, cook_reheating_2015, yadav2025reheating}.

When background metric is typically treated as classical, quantum effects of matter fields cannot be ignored in early Universe. The Semi Classical Theory of Gravity (SCTG) provides a consistent formalism in which the spacetime geometry remains classical, but matter fields are quantized. In this framework, the semiclassical Einstein equation relates the classical Einstein tensor to the expectation value of the quantum energy momentum tensor \cite{mohajan_friedmann_2013, suresh_particle_2004}. This approach has been widely employed to examine particle creation, quantum fluctuations, and nonclassical field behavior in curved spacetime \cite{kim_one-parameter_1999, finelli_quantum_1999, geralico_novel_2004, padmanabhan_gravity_2005, gangal2025dae, gangal2024particle}. It has also been shown that quantum corrections can significantly influence inflationary dynamics even when quantum gravity effects remain negligible \cite{bak1998quantum, kim1999thermal, habib1992stochastic, linde1994big}.

An essential development in this direction arises from the connection between quantum optics and cosmology. Following Kennard's foundational work on squeezed wave packets \cite{kennard_zur_1927}, squeezed and coherent squeezed states were recognized as powerful tools for analyzing nonclassical features of quantum fields in curved spacetime. Their application to FRW cosmology has yielded insights into particle production, energy density fluctuations, decoherence processes, and the statistical structure of inflaton perturbations \cite{matacz_coherent_1994, suresh_squeezed_1998, suresh_nonclassical_2001, suresh_thermal_2001, gangal2025Mandel}. Squeezed states, in particular, naturally emerge during cosmic expansion and have been extensively used to model quantum fluctuations that seed structure formation.

Recent works have extended this analysis to more refined nonclassical states such as SNS and CSNS \cite{robertson1936kinematics, shaviv_did_2011, zel1971creation, bergstrom2006cosmology, ellis1999cosmological, kuo_semiclassical_1993, caves_quantum-mechanical_1981, matacz_coherent_1994, suresh_thermal_2001, suresh_squeezed_1998, gangal2025density, suresh_nonclassical_2001, takahashi_thermo_1996, xu_quantum_2007}. These states incorporate both number state quantization and squeeze induced correlations and have been shown to exhibit distinctly nonclassical behavior in FRW cosmology. Prior studies demonstrated that SNS and CSNS significantly influence particle creation, density fluctuations, and statistical properties of the inflaton field \cite{gangal2025density}. Mandel's $Q$ parameter analysis further established their nonclassical nature in the oscillatory phase of inflation \cite{gangal2025density}.

In parallel with the cosmological applications of nonclassical states, QFT in curved spacetime has established a rigorous framework for understanding particle creation in dynamical gravitational backgrounds \cite{mahajan_particle_2008, lachieze-rey_cosmic_1995, carvalho_scalar_2004, gangal2025density, hartle_quantum_1981, gangal2024particle}. Foundational analyzes by Hawking, Parker, and Birrell Davies demonstrated that vacuum fluctuations of quantized fields become excited in space times with horizons or time dependent metrics, leading to phenomena such as Hawking radiation and cosmological particle production \cite{kennard_zur_1927, gangal2024particle, venkataratnam_particle_2004}. Within this formalism, black hole evaporation emerges from Bogoliubov transformations connecting in  and out vacua across the event horizon \cite{bakke_geometric_2009, kennard_zur_1927, venkataratnam_behavior_2013, venkataratnam_oscillatory_2010}. Although certain studies have employed thermal or thermally modified squeezed states to model aspects of Hawking emission \cite{venkataratnam_particle_2004}, such approaches remain limited to mixed state thermal ensembles. In contrast, the present work examines Hawking radiation generated from genuinely nonclassical quantum states SNS and CSNS which incorporate number-state quantization and nonclassical correlations not captured by thermal descriptions.

QFT in curved spacetime provides a fundamental framework for understanding particle creation in gravitational backgrounds, with Hawking’s discovery of black hole radiation establishing that event horizons endow spacetime with an intrinsic quantum temperature\cite{brandenberger1992entropy, sou2023decoherence}. Hawking showed that black holes emit particles due to quantum fluctuations near the horizon, producing a thermal spectrum and generating entropy in direct analogy with the laws of thermodynamics. This mechanism radically altered the classical viewpoint that black holes are perfectly absorbing objects, revealing instead that evaporation, radiation emission, and particle creation are unavoidable quantum processes \cite{gangal2025density, stoica_friedmann-lemaitre-robertson-walker_2016, hu_anisotropy_1978, dhayal_quantum_2020, venkataratnam_behavior_2013, fischetti_quantum_1979, hartle_quantum_1979}. These results introduced profound conceptual challenges, most notably the black hole information loss paradox, in which a pure state appears to evolve into a mixed thermal state, seemingly violating unitarity. Despite various attempts to address this issue including models based on dilaton gravity, lower dimensional black hole solutions, and correlation based arguments the fundamental question of information recovery remains unresolved \cite{anderson_effects_1983, suresh_nonclassical_2001, venkataratnam_density_2008, campos_semiclassical_1994, geralico_novel_2004, pedrosa_exact_2007,gangal2025density}. Entropy production associated with cosmological particle creation and gravitationally induced squeezing has therefore attracted significant renewed interest, particularly in semiclassical contexts where quantum matter fields interact with classical spacetime geometry.

Within this broader framework, nonclassical quantum states such as squeezed and coherent states have emerged as powerful tools for examining quantum processes in curved spacetime \cite{gangal2025density, robertson1936kinematics, shaviv_did_2011, zel1971creation, bergstrom2006cosmology, ellis1999cosmological, kuo_semiclassical_1993, caves_quantum-mechanical_1981, matacz_coherent_1994}. Originally developed in quantum optics, these states possess reduced quantum noise, phase sensitive correlations, and nonclassical statistical properties that make them highly suitable for studying particle creation and fluctuation phenomena in expanding universes. While some earlier studies applied thermal or thermally modified squeezed states to analyze black hole emission \cite{venkataratnam_particle_2004}, such approaches inherently describe mixed thermal ensembles. In contrast, the present work focuses on a fundamentally non thermal formulation of Hawking radiation \cite{touati2023_tunneling_noncommutative, nian2019_kerr_evaporation}. We investigate single mode emission in a flat FRW Universe when the underlying quantum field is prepared in SNS or CSNS, which encode discrete number state excitations and strong nonclassical correlations absent in thermal models \cite{robertson1936kinematics, shaviv_did_2011, zel1971creation, bergstrom2006cosmology, ellis1999cosmological, kuo_semiclassical_1993, caves_quantum-mechanical_1981, matacz_coherent_1994, suresh_thermal_2001, takahashi_thermo_1996, xu_quantum_2007, dhayal_quantum_2020, venkataratnam_oscillatory_2010, venkataratnam_nonclassical_2010}. By embedding SNS and CSNS into the semiclassical gravity framework, we derive modified expressions for Hawking flux, entropy evolution, and black hole mass variation, demonstrating that number state dependent squeezing can imprint measurable corrections on gravitational particle creation \cite{regular2024_dynamical_regular_bh, thoss2024_breakdown_evaporation}. This analysis provides a unified perspective that links cosmological nonclassicality with black hole thermodynamics and extends earlier thermal based studies to a fully non thermal, quantum state resolved description \cite{koh_gravitational_2004, lopes_gaussian_2009, lachieze-rey_theoretical_1999, sinha_[no_2003, shaviv_did_2011, berger1978classical, berger1981scalar, grishchuk1990squeezed, brandenberger1992entropy, brandenberger1993entropy,  matacz1993quantum, albrecht_inflation_1994, gasperini1993quantum, hu1994squeezed}.

The objective of this study is to construct a semiclassical formulation of single mode Hawking radiation in a spatially flat FRW Universe when the underlying quantum field is prepared in SNS or CSNS \cite{nieto1997displaced, ellis1999deviation, penzias1965measurement, handley_curvature_2021, ali2024_cr_volume_review}. By embedding these nonclassical states into the semiclassical gravity framework, we derive modified expressions for the particle spectrum, entropy, and associated evolution of black hole mass \cite{savage1986inhibition, zantedeschi2025_ultralight, aurell2024_prl_duplicate}. The analysis reveals that number state dependent squeezing imprints nontrivial corrections on Hawking emission, thereby enriching the interplay between quantum field nonclassicality and gravitational particle creation \cite{park2021_quantum_grav_bh_evolution, zhang2024_rotating_obs_signatures}. This unified treatment advances the understanding of black hole thermodynamics in cosmological settings and extends previous studies of thermal squeezed states to a fully non thermal, quantum state resolved description \cite{zhang2025_info_loss_review, corrected2023_hawking_temp}.

The paper is organized as Section 2 reviews semiclassical energy momentum tensor for a quantized inflaton in the FRW spacetime, presents the formulation of Squeezed Number States under SCTG and extends this to Coherent Squeezed Number States. In Section 3, we develop the framework for single mode Hawking radiation in an expanding FRW Universe and compute the associated entropy. Section 4 discusses the variation of black hole mass in SNS and CSNS formalisms. Finally, Secton 5 summarized the results and future work directions.

\section{Semiclassical Framework for Nonclassical Inflaton, Energy Momentum Tensor and Foundations of SNS/CSNS}

A wide class of contemporary cosmological models is constructed by combining Einstein’s classical gravitational field equations with a quantized scalar field evolving in an FRW background. In this semiclassical approach, the spacetime geometry remains classical, while matter is treated quantum mechanically an approximation known as SCTG \cite{gangal2024particle}. Adopting natural units $c=\hbar=1$ and $\mathcal{G} = 1/m_{p}^{2}$, the SCEE takes the form \cite{gangal2025density, gangal2024particle, gangal2025Mandel}  
\begin{equation} \label{2.1}
\mathcal{E}_{\mu \nu }=\frac{8\pi\left\langle :\mathcal{T}_{\mu \nu }:\right\rangle}{m_{p}^{2}},
\end{equation}
where $\left\langle :\mathcal{T}_{\mu \nu }:\right\rangle$ denotes normal   ordered expectation value of the energy   momentum tensor and $\mathcal{E}_{\mu \nu}$ is the Einstein tensor associated with the FRW geometry. The $|\psi\rangle$ describes the inflaton satisfies the time dependent Schrödinger equation  
\begin{equation} \label{2.2}
 \hat{\mathcal{H}}\,|\psi\rangle=i\,\frac{\partial}{\partial t}|\psi\rangle,
\end{equation}
with $ \hat{\mathcal{H}}$ representing the Hamiltonian operator of the quantized field. In the flat spatial FRW spacetime, the generalized coordinates $(r_{1},r_{2},r_{3},r_{4})$ obey  
\begin{equation} \label{2.3}
\left(dr_{1}^{2}+dr_{2}^{2}+dr_{3}^{2}\right)\mathcal{G}^{2}(t)-dr_{4}^{2}=ds^{2},
\end{equation}
where $\mathcal{G}(t)$ is the cosmological scale factor. For a homogeneous inflaton field $\Phi(t)$ \cite{gangal2025density, gangal2024particle, gangal2025Mandel}, the Lagrangian density  
\begin{equation} \label{2.4}
\mathfrak{L}=-\frac{\sqrt{-\mathfrak{g}}\left(m^{2}\Phi^{2}+\mathfrak{g}^{\mu\nu}\partial_{\mu}\Phi\,\partial_{\nu}\Phi\right)}{2},
\end{equation}
reduces, using the metric \eqref{2.3}, to  
\begin{equation} \label{2.5}
\mathfrak{L}=\frac{\mathcal{G}^{3}(t)}{2}\left(\dot{\Phi}^{2}-m^{2}\Phi^{2}\right).
\end{equation}

The corresponding Klein Gordon equation is obtained as \cite{kennard_zur_1927, gangal2025density, dhayal_quantum_2020, gangal2025Mandel}  
\begin{equation} \label{2.6}
\ddot{\Phi}+\frac{3\dot{\mathcal{G}}(t)}{\mathcal{G}(t)}\dot{\Phi}+m^{2}\Phi=0.
\end{equation}

Under canonical quantization, the Hamiltonian of the inflaton field yields the expectation value \cite{gangal2025Mandel}  
\begin{equation}  \label{2.8}
\langle : \hat{\mathcal{H}}_{m}:\rangle
=\frac{m^{2}\mathcal{G}^{3}(t)}{2}\left\langle : \hat{\Phi}^{2}:\right\rangle
+\frac{1}{2\mathcal{G}^{3}(t)}\left\langle : \hat{\Pi}^{2}:\right\rangle,
\end{equation}
The temporal component of energy momentum tensor \cite{gangal2025density, gangal2024particle, gangal2025Mandel} is then  
\begin{equation} \label{2.9}
\mathcal{T}_{00}=\mathcal{G}^{3}(t)\left(\frac{m^{2} \hat{\Phi}^{2}}{2}+\frac{\dot{\Phi}^{2}}{2}\right).
\end{equation}

Single mode squeezed state is constructed by successive action of squeezing $ \hat{W}(\rho ,\Psi)$ and displacement operators $\mathfrak{D}(\Upsilon)$ on the vacuum, and is defined as  
\begin{equation} \label{3.1}
|\Upsilon ,\zeta \rangle= \hat{W}(\rho ,\Psi)\,\mathfrak{D}(\Upsilon)\,|0\rangle,
\end{equation}
where $\mathfrak{D}(\Upsilon)$ and $ \hat{W}(\rho ,\Psi)$ are given by  
\begin{equation} \label{3.2}
\mathfrak{D}(\Upsilon)=\exp\!\left(\Upsilon\, \hat{e}^{\dagger}-\Upsilon^{*} \hat{e}\right),
\end{equation}
\begin{equation} \label{3.3}
 \hat{W}(\rho ,\Psi)=\exp\!\left[\frac{1}{2}\!\left(\rho\,e^{-i\Psi} \hat{e}^{2}-\rho\,e^{i\Psi} \hat{e}^{\dagger 2}\right)\right].
\end{equation}
Here, the squeezing parameter $\rho$ ranges from $0\leq\rho<\infty$ and the squeezing angle $\Psi$ satisfies $-\pi\leq\Psi\leq\pi$. The operator $ \hat{W}(\rho,\Psi)$ transforms the mode operators according to  
\begin{equation} \label{3.4}
 \hat{W}^{\dagger}\, \hat{e}\, \hat{W}
=\cosh\rho\, \hat{e}-e^{i\Psi}\sinh\rho\, \hat{e}^{\dagger},
\end{equation}
\begin{equation} \label{3.5}
 \hat{W}^{\dagger}\, \hat{e}^{\dagger}\, \hat{W}
=\cosh\rho\, \hat{e}^{\dagger}-e^{-i\Psi}\sinh\rho\, \hat{e}.
\end{equation}

Applying squeezing operator to a number state produces a SNS,
\begin{equation} \label{3.6}
 \hat{W}(\rho,\Psi)\,|n\rangle = |\zeta ,n\rangle.
\end{equation}

The annihilation and creation operators $ \hat{e}$ and $ \hat{e}^{\dagger}$ act on the number state basis as \cite{kim2000nonequilibrium}  
\begin{equation} \label{3.7}
 \hat{e}\,|n, t\rangle=\sqrt{n}\,|n-1, t\rangle,
\end{equation}
\begin{equation} \label{3.8}
 \hat{e}^{\dagger}\,|n, t\rangle=\sqrt{n+1}\,|n+1, t\rangle,
\end{equation}
with commutation relation as  
\begin{equation} \label{3.10}
[ \hat{e}, \hat{e}^{\dagger}]=1.
\end{equation}
The corresponding number operator satisfies  
\begin{equation} \label{3.11}
 \hat{e}^{\dagger} \hat{e}(t)\,|n, t\rangle = n\,|n, t\rangle.
\end{equation}

We now describe the displacement operation associated with the construction of CSNS \cite{ venkataratnam_density_2008, venkataratnam_oscillatory_2010, venkataratnam_behavior_2013}. The single mode coherent state is generated by applying $\mathfrak{D}(\Upsilon)$ to vacuum,
\begin{equation} \label{4.1}
\mathfrak{D}(\Upsilon)\,|0\rangle = |\Upsilon\rangle,
\end{equation}
and satisfies the eigenvalue relation  
\begin{equation} \label{4.2}
 \hat{e}\,|\Upsilon\rangle = \Upsilon\,|\Upsilon\rangle.
\end{equation}
The displacement operator transforms the ladder operators according to
\begin{equation} \label{3.18}
\mathfrak{D}^{\dagger}(\Upsilon)\, \hat{e}^{\dagger}\,\mathfrak{D}(\Upsilon)
   =  \hat{e}^{\dagger}+\Upsilon^{*},
\end{equation}
\begin{equation} \label{3.19}
\mathfrak{D}^{\dagger}(\Upsilon)\, \hat{e}\,\mathfrak{D}(\Upsilon)
   =  \hat{e}+\Upsilon.
\end{equation}

Coherent squeezed vacuum state is obtained by the joint action of the displacement and squeezing operators on $|0\rangle$,
\begin{equation} \label{4.5}
\mathfrak{D}(\Upsilon)\, \hat{W}(\rho,\Psi)\,|0\rangle
   = |\Upsilon,\zeta,0\rangle.
\end{equation}
Similarly, acting on a number state yields CSNS,
\begin{equation} \label{4.6}
\mathfrak{D}(\Upsilon)\, \hat{W}(\rho,\Psi)\,|n\rangle
   = |\Upsilon,\zeta,n\rangle.
\end{equation}

\section{Single Mode Hawking's Radiation in Flat FRW Universe For Nonclassical SNS and CSNS}

Black holes are regions of space time where the gravitational field becomes strong that no classical particle/radiation can escape once it crosses event horizon \cite{kumar2024_pure_states_hawking, ravuri2025_hawking_charged_cosmological}. Within the framework of General Relativity (GR), their classical description originates from exact solutions of the EFE, such as the Schwarzschild and Kerr metrics \cite{ venkataratnam_density_2008, venkataratnam_oscillatory_2010, venkataratnam_behavior_2013}. However, the thermal radiation predicted by Hawking cannot be accounted for by classical GR \cite{anderson_effects_1983, campos_semiclassical_1994}. Hawking demonstrated in his seminal work that quantum field fluctuations near the event horizon lead to spontaneous particle creation, giving rise to a thermal spectrum \cite{venkataratnam_density_2008, kumar2024_pure_states_hawking, ravuri2025_hawking_charged_cosmological, venkataratnam_behavior_2013, venkataratnam_nonclassical_2010, venkataratnam_density_2008, venkataratnam_nonclassical_2010, venkataratnam_particle_2004}. For the extremal Kerr black holes, it is generally accepted that the radiation vanishes in the extremal limit, illustrating the subtle interplay between angular momentum and horizon structure.

The existence of Hawking radiation fundamentally alters the classical picture of black holes. The associated thermal emission induces mass loss, entropy generation, and establishes a deep connection with black hole thermodynamics \cite{delrio2025_backreaction_review}. These processes play a significant role in contemporary cosmology and high energy astrophysics. Nevertheless, the transition from an initial pure state to a mixed thermal state conflicts with quantum mechanical unitarity, leading to the long standing \textquotedblleft information loss paradox\textquotedblright\. Several approaches have been proposed to address this issue, including models based on two dimensional dilaton gravity  and $(2+1)$ dimensional anti de Sitter black holes \cite{akers2024_relative_state_counting}. Despite these efforts, a complete resolution remains elusive. Hawking's perspective suggests that information loss is an inherent consequence of combining classical gravity with quantum mechanics. However, when off diagonal correlations are preserved, the density matrix remains pure, consistent with analyses demonstrating that quantum correlations among subsystems encode the full information content. Entropy generation in cosmological particle creation has also been extensively discussed in the literature \cite{aurell2024_prl_duplicate}.

In this study, motivated by these developments, we investigate Hawking's radiation in the framework of single mode nonclassical number state formalisms within a flat FRW universe. We utilize nonclassical states characterized by discrete occupation and enhanced quantum correlations, specifically for SNS and CSNS \cite{ellis1999cosmological, thoss2024_breakdown_evaporation}. These states provide a refined description of particle creation, entropy flow, and horizon thermodynamics, allowing us to analyze the dependence of Hawking radiation on the squeezing parameter $\rho$, occupation number $n$, coherent displacement ${\varUpsilon}$, and the background FRW expansion.

Hawking's effect describes creation of particle antiparticle pairs in the vicinity of a horizon, where one member escapes to infinity while the other falls into the black hole. The mean number of particles in any specific mode is conventionally given by

\begin{equation} \label{5.1}
\langle \hat{n} \rangle = \frac{1}{e^{\omega/{\mathbb{T}}_{\mathbb{H}}} - 1},
\end{equation}

where ${\mathbb{T}}_{\mathbb{H}}$ is called as Hawking temperature. In the present analysis, we computed the ${\mathbb{T}}_{\mathbb{H}}$ for SNS and CSNS. These corrections generalize the Hawking spectrum to cases where quantum state of the field deviates from the standard thermal or vacuum configurations. So the mean number of particles radiated to a particular mode can be written by considering $ |0> $ as the vacuum state, then   equation (\ref{5.1}) can be written as 

\begin{equation}\label{5.2}
\langle0|\hat{n}|0\rangle = \frac{1}{e^{\omega/{\mathbb{T}}_{\mathbb{H}}} - 1} ,
\end{equation}

where $ \hat{n} $ is number operator. Using equation (\ref{3.6}) (the operation of squeezing operator to a number state) and (\ref{5.2}), the  spectrum of black body radiation in single mode SNS can be written as

\begin{equation}\label{5.3}
\langle\zeta ,n|\hat{n}|\zeta ,n\rangle_{SNS} = \frac{1}{e^{\omega/{\mathbb{T}}_{\mathbb{H}}} - 1},
\end{equation}

Using Eqs. (\ref{3.4}-\ref{3.5}, \ref{3.11}) the values of $\langle\zeta ,n|\hat{n}|\zeta ,n\rangle_{SNS}$

\begin{equation}\label{5.4}
\langle\zeta ,n|\hat{n}|\zeta ,n\rangle_{SNS} = \langle n|\overset{\wedge}W^{\dagger}(\rho ,\Psi )\overset{\wedge }{\mathit{e}}^{\dagger }\overset{\wedge }{\mathit{e}}\overset{\wedge}W(\rho ,\Psi )|n\rangle,
\end{equation}

or 

\begin{align}\label{5.5}
\langle\zeta ,n|\hat{n}|\zeta ,n\rangle_{SNS} = &\langle n|[(\text{cosh$\rho
$})(\overset{\wedge }{\mathit{e}}^{\dagger }) - (\exp (-\mathit{i}\Psi )\text{sinh$\rho $})(\overset{\wedge }{\mathit{e}})][(\text{cosh$\rho $})(\overset{\wedge }{\mathit{e}})  \nonumber\\
&- (\exp (\mathit{i}\Psi )\text{sinh$\rho $})(\overset{\wedge
}{\mathit{e}} ^{\dagger })]|n\rangle,
\end{align}

using Eqs. (\ref{3.7}-\ref{3.10}) in Eq. (\ref{5.5}), normal order expectation value of number operator in cosmological reference can be computed as

\begin{equation}\label{5.6}
\langle\zeta ,n|\hat{n}|\zeta ,n\rangle_{SNS} = (n)\text{Cosh}(2\rho)+\text{Sinh}^2\rho,
\end{equation}

comparing Eq. (\ref{5.3}) and Eq. (\ref{5.6}) we get

\begin{equation}\label{5.7}
\langle\zeta ,n|\hat{n}|\zeta ,n\rangle_{SNS} = \frac{1}{e^{\omega/{\mathbb{T}}_{\mathbb{H}}} - 1}=(n)\text{Cosh}(2\rho)+\text{Sinh}^2\rho,
\end{equation}

so Hawking temperature ${\mathbb{T}}_{{\mathbb{H}}_{SNS}}$ for single mode SNS can be written as

\begin{equation}\label{5.8}
 {\mathbb{T}}_{{\mathbb{H}}_{SNS}} ={\omega\over\ln \left(1+{1\over\ ((n)\text{Cosh}(2\rho)+\text{Sinh}^2\rho)}\right)}.
\end{equation}

\begin{figure}
\begin{center} 
\includegraphics[width=0.68\textwidth]{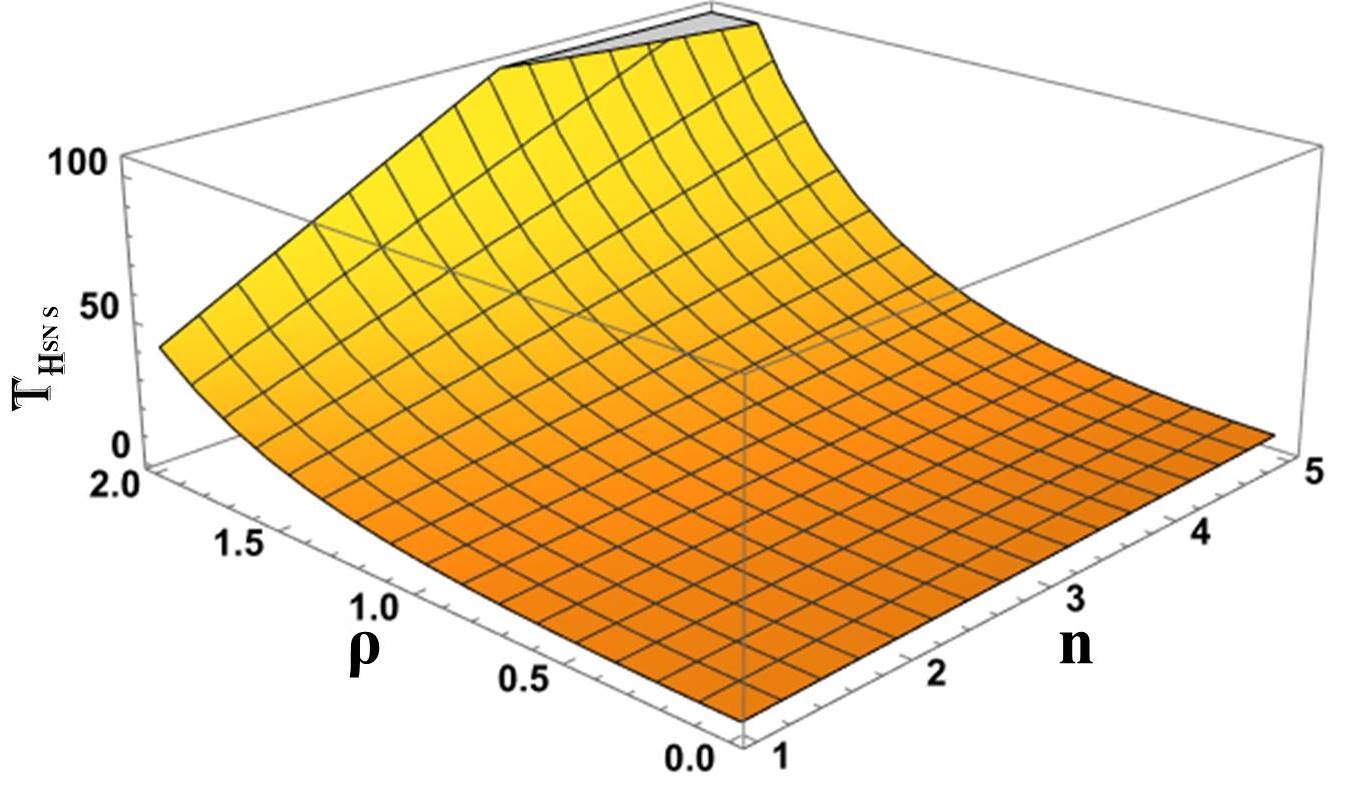}
\caption{3D plot for Hawking temperature $ {\mathbb{T}}_{{\mathbb{H}}_{SNS}} $ with $n$ and $ \rho $}
\label{fig:figure_1}
 \end{center}
 \end{figure}
 
\begin{figure}
\begin{center} 
\includegraphics[width=0.68\textwidth]{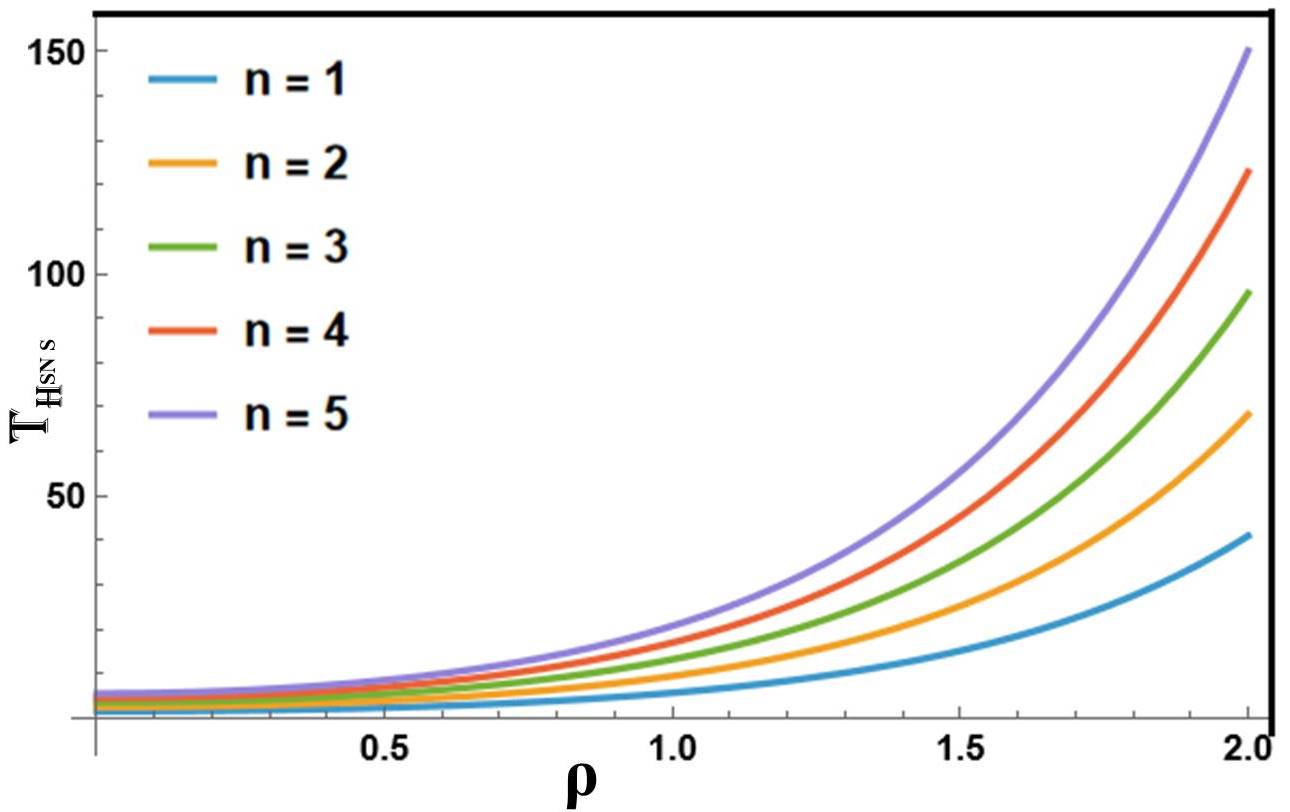}
\caption{2D plot for Hawking temperature $ {\mathbb{T}}_{{\mathbb{H}}_{SNS}} $ with  $ \rho $, for different values of $n$. }
\label{fig:figure_2}
\end{center}
\end{figure}

Similarly, Using Eqs. (\ref{4.5} and \ref{4.6}) (the operation of displacement operator $\mathfrak{D}$($\Upsilon $) and the squeezing operator \(\overset{\wedge}W(\rho ,\Psi )\) to a number state) and \ref{5.2}, spectrum of black body radiation in single mode CSNS can be written as

\begin{equation}\label{5.9}
\langle\Upsilon ,\zeta ,n|\hat{n}|\Upsilon ,\zeta ,n\rangle_{CSNS} = \frac{1}{e^{\omega/{\mathbb{T}}_{\mathbb{H}}} - 1},
\end{equation}

Using Eqs. (\ref{3.4}-\ref{3.5}, \ref{3.11}, \ref{4.5}-\ref{4.6}) the values of $\langle\Upsilon ,\zeta ,n|\hat{n}|\Upsilon ,\zeta ,n\rangle_{CSNS}$

\begin{equation}\label{5.10}
\langle\Upsilon ,\zeta ,n|\hat{n}|\Upsilon ,\zeta ,n\rangle_{CSNS} = \langle n|\overset{\wedge}W ^{\dagger }(\rho ,\Psi )\mathfrak{D}^{\dagger }(\Upsilon)\overset{\wedge }{\mathit{e}}^{\dagger }\overset{\wedge }{\mathit{e}}\mathfrak{D}(\Upsilon)\overset{\wedge}W(\rho ,\Psi )|n\rangle,
\end{equation}

or

\begin{equation}\label{5.11}
\langle\Upsilon ,\zeta ,n|\hat{n}|\Upsilon ,\zeta ,n\rangle_{CSNS} = \langle n|\overset{\wedge}W ^{\dagger }(\rho ,\Psi )(\overset{\wedge }{\mathit{e}} ^{\dagger }+\Upsilon ^*)(\overset{\wedge }{\mathit{e}} +\Upsilon)\overset{\wedge}W(\rho ,\Psi )|n\rangle,
\end{equation}

using Eqs. (\ref{3.7}-\ref{3.10}) in Eq. (\ref{5.11}), normal order expectation value of number operator in cosmological reference can be computed as

\begin{equation}\label{5.12}
\langle\Upsilon ,\zeta ,n|\hat{n}|\Upsilon ,\zeta ,n\rangle_{CSNS} = {\varUpsilon ^2}+(n)\text{Cosh}(2\rho)+\text{Sinh}^2\rho,
\end{equation}

comparing Eq. (\ref{5.9}) and Eq. (\ref{5.12}) we get

\begin{equation}\label{5.13}
\langle\Upsilon ,\zeta ,n|\hat{n}|\Upsilon ,\zeta ,n\rangle_{CSNS} = \frac{1}{e^{\omega/{\mathbb{T}}_{\mathbb{H}}} - 1}={\varUpsilon ^2}+(n)\text{Cosh}(2\rho)+\text{Sinh}^2\rho,
\end{equation}

so Hawking temperature ${\mathbb{T}}_{{\mathbb{H}}_{CSNS}}$ for single mode CSNS can be written as

\begin{equation}\label{5.14}
 {\mathbb{T}}_{{\mathbb{H}}_{CSNS}} ={\omega\over\ln \left(1+{1\over\ ({\varUpsilon ^2}+(n)\text{Cosh}(2\rho)+\text{Sinh}^2\rho)}\right)},
\end{equation}

\begin{figure}
\begin{center} 
\includegraphics[width=0.68\textwidth]{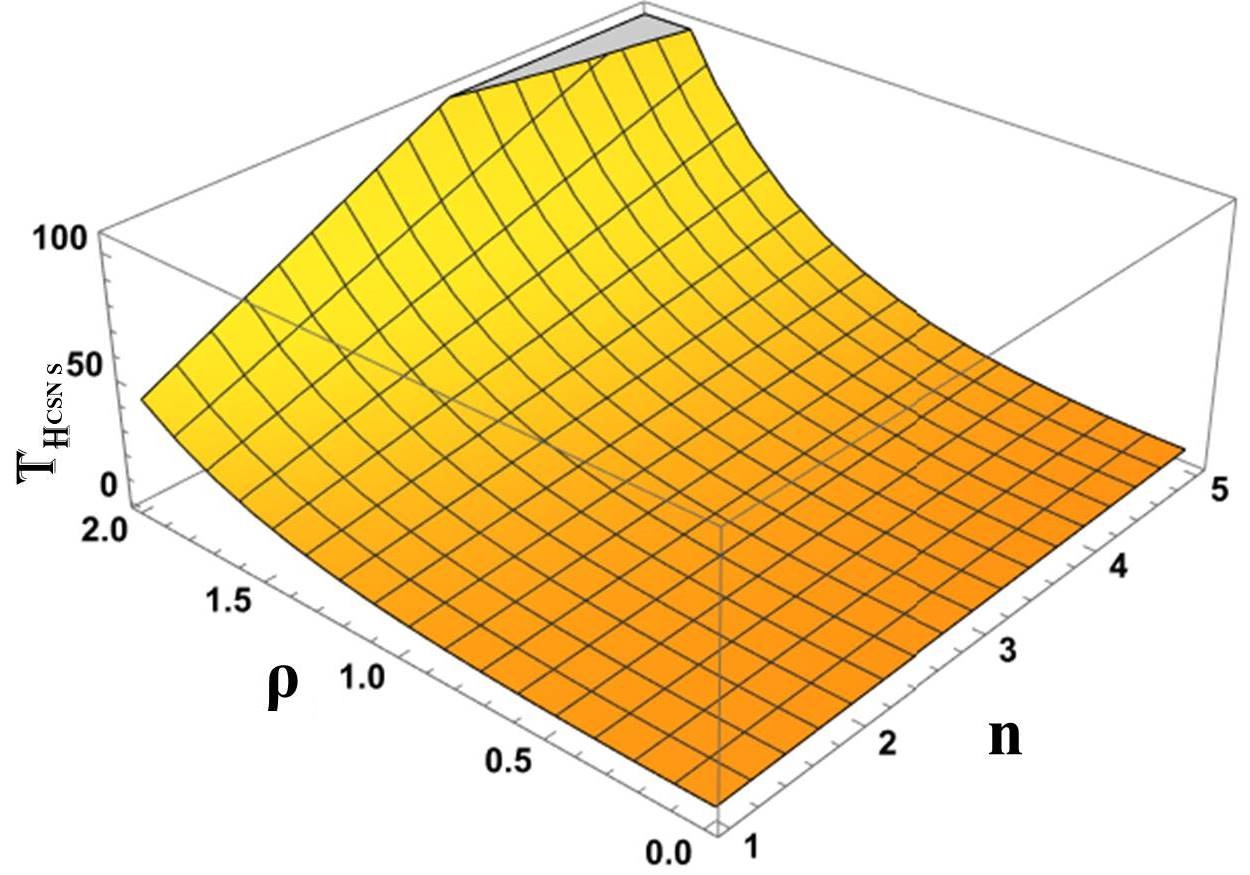}
\caption{3D plot for Hawking temperature $ {\mathbb{T}}_{{\mathbb{H}}_{CSNS}} $ with $n$ and $ \rho $}
\label{fig:figure_3}
 \end{center}
 \end{figure}
 
\begin{figure}
\begin{center} 
\includegraphics[width=0.68\textwidth]{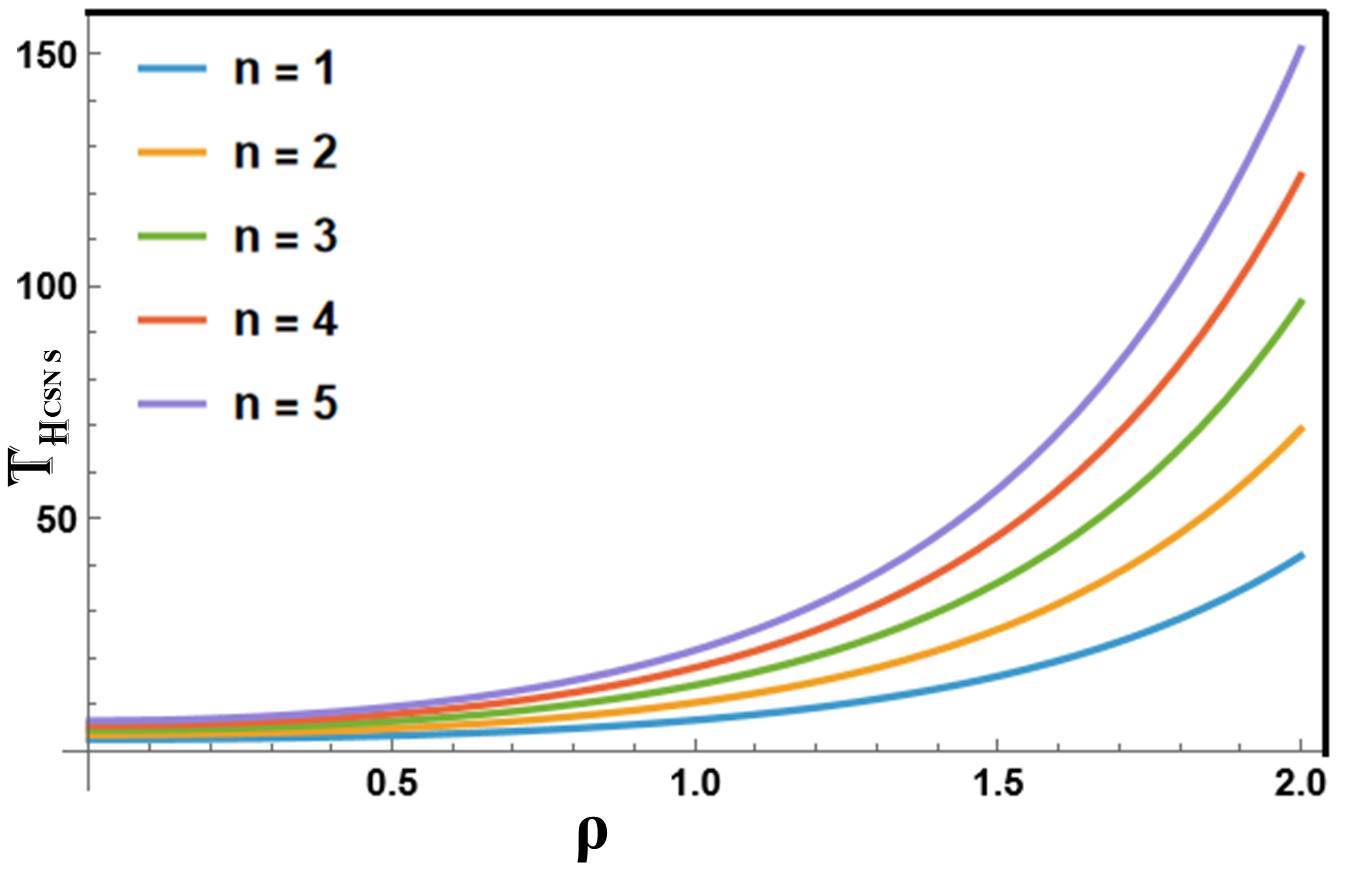}
\caption{2D plot for Hawking temperature $ {\mathbb{T}}_{{\mathbb{H}}_{CSNS}} $ with $ \rho $, for different values of $n$. }
\label{fig:figure_4}
\end{center}
\end{figure}

\section{Entropy Formulation: Flat FRW Universe For Single Mode  Nonclassical SNS and CSNS}

Entropy generation in expanding FRW backgrounds offers a crucial window into the quantum structure of horizon induced particle creation. For nonclassical SNS and CSNS, entropy encodes squeezing driven correlations and deviations from thermal behavior, providing insights into information flow across cosmological horizons. The relation between a black hole's entropy and Hawking's temperature can be expressed as: 
 
\begin{equation}\label{6.1}
{\mathbb{T}}_{\mathbb{H}}=\left[\left({\Delta {\mathbb{S}}\over\Delta
{\mathbb{M}}}\right)_{Q,J}\right]^{-1}\mathbb{G}^2=\left[\left({\partial {\mathbb{S}}\over\partial
{\mathbb{M}}}\right)_{Q,J}\right]^{-1}\mathbb{G}^2,
\end{equation}

where ${\mathbb{S}}$ is entropy,  ${\mathbb{M}}$ is mass parameter, $Q$ is charge and $J$ is angular momentum. Now, we simplify Eq.~(\ref{6.1}) and obtain 

\begin{equation}\label{6.2}
\Delta {\mathbb{S}} = {\mathbb{G}^2 \over {\mathbb{T}}_{\mathbb{H}}}\,\Delta {\mathbb{M}},
\end{equation}

where $\Delta {\mathbb{M}}$ is mass variations and $\Delta {\mathbb{S}}$ is variations in entropy. Hence, Eq.~(\ref{6.2}) expresses the entropy change at given ${\mathbb{T}}_{\mathbb{H}}$ in terms of ${\mathbb{M}}$.  

In this section, using the single mode number state formalism, we compute the entropy variation associated with black hole evaporation. Now as per Shannon’s information theory, entropy is defined as 

\begin{equation}\label{6.3}
{\mathbb{S}} = -{\rm Tr}\left({\mathcal{P}} \ln {\mathcal{P}}\right)=-\int ~{\mathcal{P}} \ln{\mathcal{P}} d^2 z,
\end{equation}

where ${\mathcal{P}}$ is called density matrix. For ${\rm Tr}({\mathcal{P}}^{2})=1$, system remains pure state; if ${\rm Tr}({\mathcal{P}}^{2})<1$, it is mixed, with the limiting maximally mixed case satisfying ${\rm Tr}({\mathcal{P}}^{2})=0$, implying the absence of retrievable information. The representation of Eq.~(\ref{6.3}) in the single mode SNS framework can be written as

\begin{equation}\label{6.4}
{\mathbb{S}} = -Tr ~({\mathcal{P}}_{SNS} \ln{\mathcal{P}}_{SNS})= -\int ~{\mathcal{P}}_{SNS} \ln{\mathcal{P}}_{SNS} d^2 z,
\end{equation}

Similarly, the representation of Eq.~(\ref{6.3}) in the single mode CSNS framework can be written as

\begin{equation}\label{6.5}
{\mathbb{S}} = -Tr ~({\mathcal{P}}_{CSNS} \ln{\mathcal{P}}_{CSNS})= -\int ~{\mathcal{P}}_{CSNS} \ln{\mathcal{P}}_{CSNS} d^2 z,
\end{equation}

where in Eq.~(\ref{6.3})

\begin{equation}\label{6.6}
\mathrm{Tr}\,{\mathcal{P}} 
= 2\pi \int d^{2}\varUpsilon\ d^{2}\zeta \,
|\sigma(\varUpsilon,\zeta)|^{2}
\,|\varUpsilon,\zeta\rangle \langle \varUpsilon',\zeta'| \,
|\varUpsilon,\zeta\rangle \langle \varUpsilon',\zeta'| 
= 1,
\end{equation}

and

\begin{equation}\label{6.7}
{\mathcal{P}} = |\varUpsilon,\zeta\rangle \langle \varUpsilon,\zeta|
= \int d^{2}\varUpsilon\, d^{2}\zeta \,
|\sigma(\varUpsilon, \zeta)|^{2}
\,|\varUpsilon,\zeta\rangle \langle \varUpsilon,\zeta|,
\end{equation}

where the random phase approximation replaces

\begin{equation}\label{6.8}
\int d^{2}\varUpsilon\, d^{2}\zeta 
\int d^{2}\varUpsilon'\, d^{2}\zeta'\,
\sigma^{*}(\varUpsilon, \zeta)\sigma(\varUpsilon, \zeta)
|\varUpsilon,\zeta\rangle \langle \varUpsilon,\zeta|,
\end{equation}

with

\begin{equation}\label{6.9}
\int d^{2}\zeta \int d^{2}(\varUpsilon \zeta)'
\,\delta(\varUpsilon' \zeta' - \varUpsilon \zeta)\,
\sigma(\varUpsilon, \zeta)\,
|\varUpsilon,\zeta\rangle \langle \varUpsilon',\zeta'|.
\end{equation}

$|\sigma(\varUpsilon\, \zeta)|^{2}|\varUpsilon,\zeta\rangle \langle \varUpsilon,\zeta|$ is diagonal elements of density matrix, so for each value of $(\varUpsilon,\zeta)$, so the appropriate Gaussian distribution is written as

\begin{equation}\label{6.10}
{\mathcal{P}} (z) = {1\over\pi\langle\hat{n}\rangle}\exp\left[-{|z|^2\over\langle\hat{n}\rangle}\right].
\end{equation}

or

\begin{equation}\label{6.11}
\ln {\mathcal{P}} (z) = -\ln (\pi\langle\hat{n}\rangle)-{|z|^2\over\langle\hat{n}\rangle}.
\end{equation}

using Eq. (\ref{6.11}) in Eq. (\ref{6.3}) 

\begin{equation}\label{6.12}
{\mathbb{S}} = -\int ~{\mathcal{P}} (z) \left[ -\ln (\pi\langle\hat{n}\rangle)-{|z|^2\over\langle\hat{n}\rangle}\right] d^2 z,
\end{equation}

here the ${\mathcal{P}}$ is normalized then $\int{\mathcal{P}}(z)d^2 z=1$ and $\int z^2{\mathcal{P}}(z)d^2 z=\langle\hat{n}\rangle$, using them in Eq. (\ref{6.12}) entropy will be  

\begin{equation}\label{6.13}
{\mathbb{S}} = 1+\ln (\pi\langle\hat{n}\rangle)=1+ \ln \langle\hat{n}\rangle+\ln (\pi) ,
\end{equation}

for single mode SNS, using Eqs. (\ref{5.2}-\ref{5.3}) the Eq. (\ref{6.13})  can be rewritten as  

\begin{equation}\label{6.14}
{\mathbb{S}}_{SNS} = 1+\ln (\pi\langle\hat{n}\rangle)=1+ \ln \langle\zeta ,n|\hat{n}|\zeta ,n\rangle_{SNS}+\ln (\pi) ,
\end{equation}

here using the Eq. (\ref{5.7}) in Eq. (\ref{6.14})

\begin{equation}\label{6.15}
{\mathbb{S}}_{SNS} = 1+\ln (\pi\langle\hat{n}\rangle)=1+\ln (\pi) + \ln [(n)\text{Cosh}(2\rho)+\text{Sinh}^2\rho)],
\end{equation}

change in entropy can be defined as 

\begin{equation}\label{6.16}
\Delta {\mathbb{S}} = {\mathbb{S}}(\rho)-{\mathbb{S}} (0),
\end{equation}

here using the Eq. (\ref{6.15}) in Eq. (\ref{6.16})

\begin{equation}\label{6.17}
\Delta {\mathbb{S}}_{SNS} = \ln \left[{(n)\text{Cosh}(2\rho)+\text{Sinh}^2\rho)}\over n \right],
\end{equation}

\begin{figure}[htb]
\begin{center} 
\includegraphics[width=0.68\textwidth]{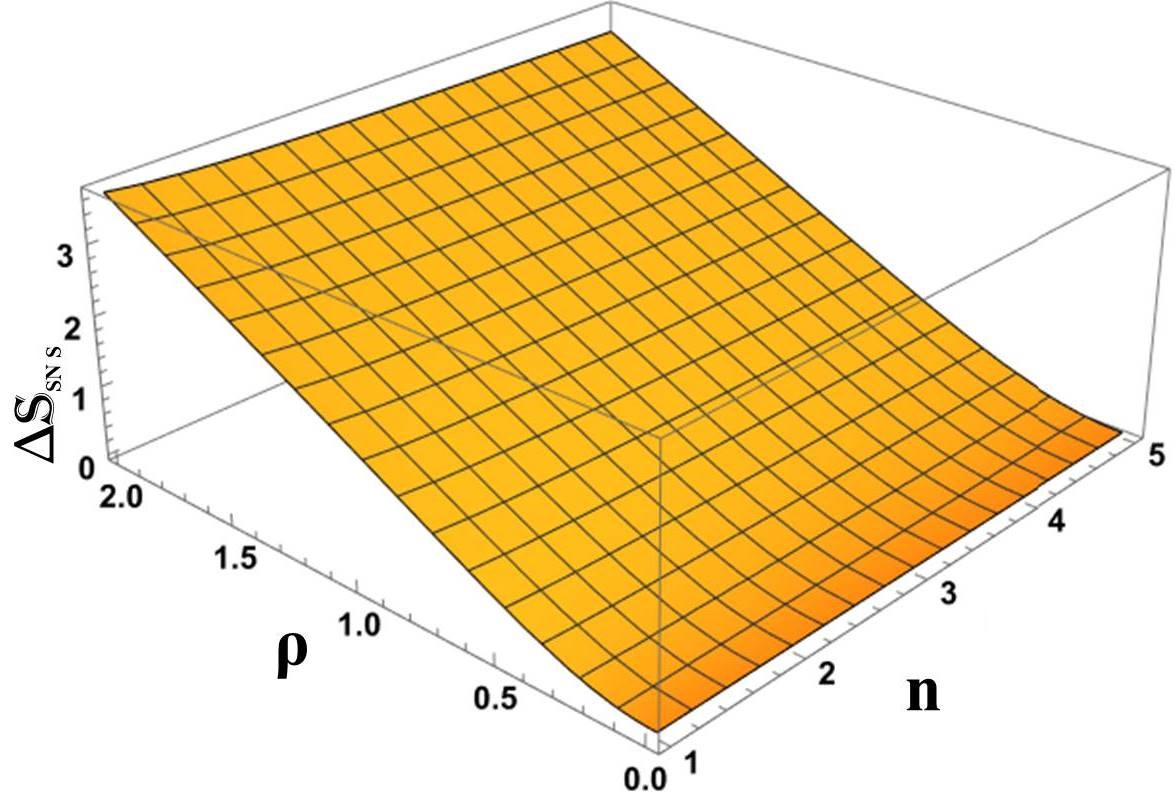}
\caption{3D Plot for $ \Delta {\mathbb{S}}_{SNS} $ with number state parameter ($ n $) and squeezing parameter ($ \rho $).}
\label{fig:figure_5}
\end{center}
\end{figure}

\begin{figure}[htb]
\begin{center} 
\includegraphics[width=0.68\textwidth]{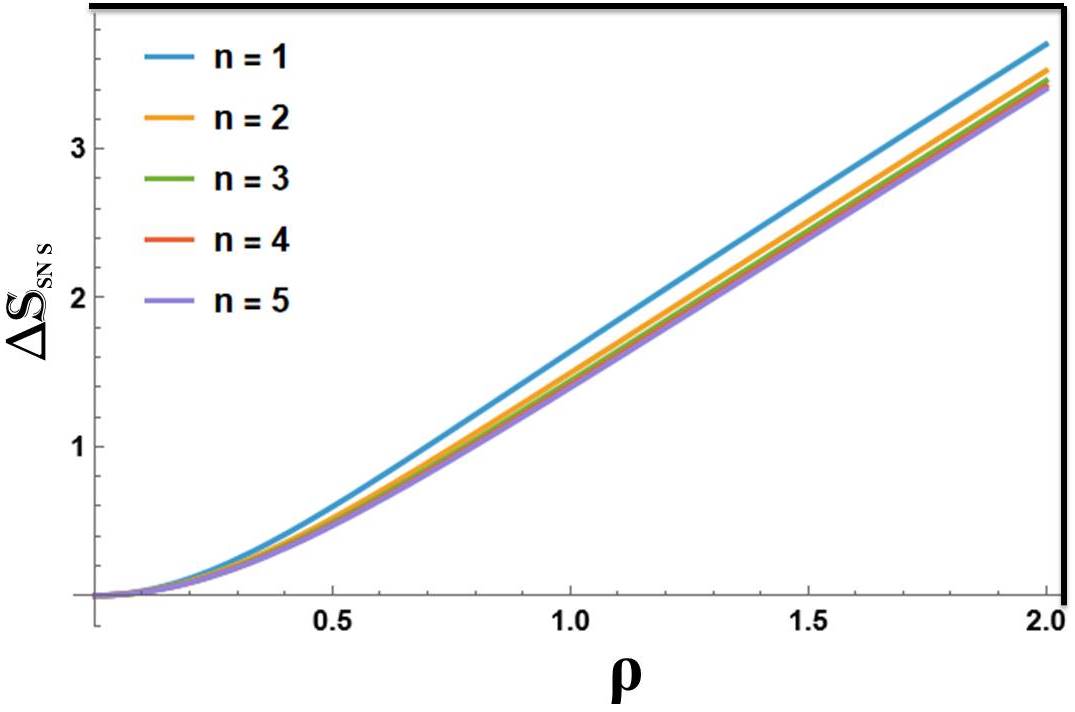}
\caption{2D Plot for $ \Delta {\mathbb{S}}_{SNS} $ and squeezing parameter ($ \rho $) for different values of number state parameter (n).}
\label{fig:figure_6}
\end{center}
\end{figure}
 
now using the Eq. (\ref{6.2})

\begin{equation}\label{6.18}
\Delta {\mathbb{M}} = {{\mathbb{T}}_{\mathbb{H}}\over \mathbb{G}^2}~\Delta {\mathbb{S}},
\end{equation}

the variation in mass $\Delta {\mathbb{M}}_{SNS}$ of the black hole for for single mode SNS, using Eqs. (\ref{5.8}, \ref{6.17}) in the Eq. (\ref{6.18})

\begin{equation}\label{6.19}
\Delta {\mathbb{M}}_{SNS} = {{\omega\over \mathbb{G}^2 \ln \left(1+{1\over\ ((n)\text{Cosh}(2\rho)+\text{Sinh}^2\rho)}\right)}}~\ln \left[{(n)\text{Cosh}(2\rho)+\text{Sinh}^2\rho)}\over n \right],
\end{equation}

\begin{figure}[htb]
\begin{center} 
\includegraphics[width=0.68\textwidth]{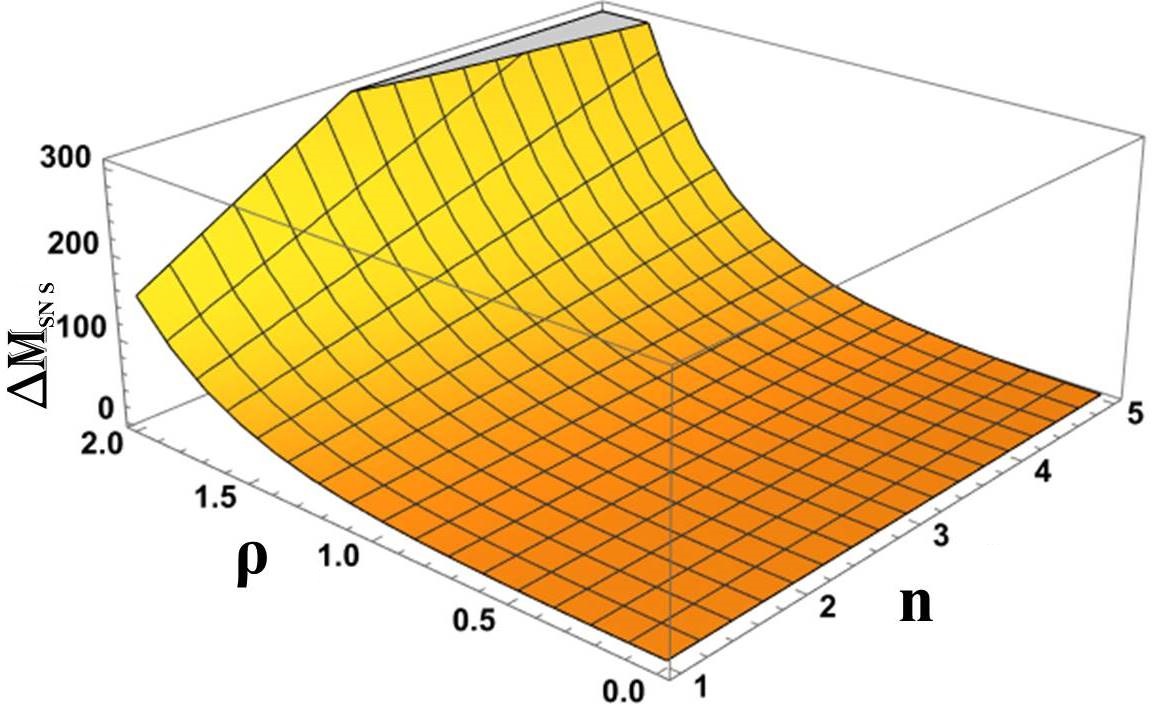}
\caption{3D Plot for $ \Delta {\mathbb{M}}_{SNS} $ with number state parameter $ n $ and squeezing parameter $ \rho $.}
\label{fig:figure_7}
\end{center}
\end{figure}

\begin{figure}[htb]
\begin{center} 
\includegraphics[width=0.68\textwidth]{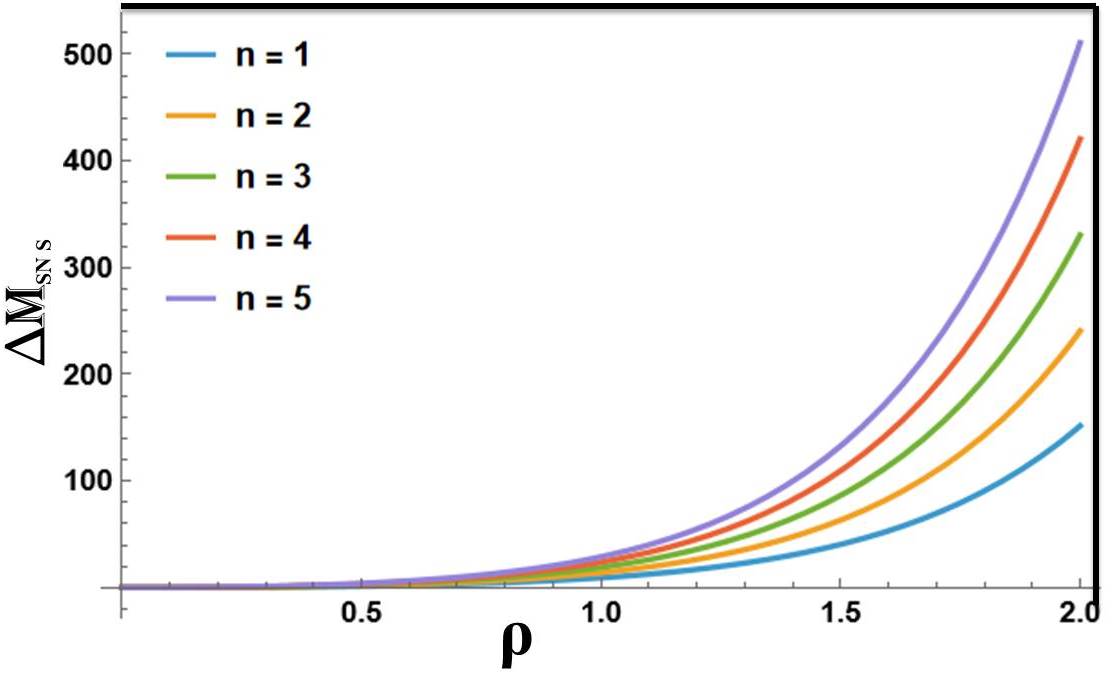}
\caption{2D Plot for $ \Delta {\mathbb{M}}_{SNS} $ and squeezing parameter ($ \rho $) for different values of number state parameter (n).}
\label{fig:figure_8}
\end{center}
\end{figure}

similarly for single mode CSNS, using Eqs. (\ref{5.2}, \ref{5.9}) the Eq. (\ref{6.13})  can be rewritten as

\begin{equation}\label{6.20}
{\mathbb{S}}_{CSNS} = 1+\ln (\pi) + \ln \langle\Upsilon ,\zeta ,n|\hat{n}|\Upsilon ,\zeta ,n\rangle_{CSNS},
\end{equation}

here using the Eq. (\ref{5.12}) in Eq. (\ref{6.20})

\begin{equation}\label{6.21}
{\mathbb{S}}_{CSNS} = 1+\ln (\pi) + \ln [{\varUpsilon ^2}+ (n)\text{Cosh}(2\rho)+\text{Sinh}^2\rho)],
\end{equation}

here using the Eq. (\ref{6.21}) in Eq. (\ref{6.16})

\begin{equation}\label{6.22}
\Delta {\mathbb{S}}_{CSNS} = \ln \left[{\varUpsilon ^2}+{(n)\text{Cosh}(2\rho)+\text{Sinh}^2\rho)}\over n \right],
\end{equation}

\begin{figure}[htb]
\begin{center} 
\includegraphics[width=0.68\textwidth]{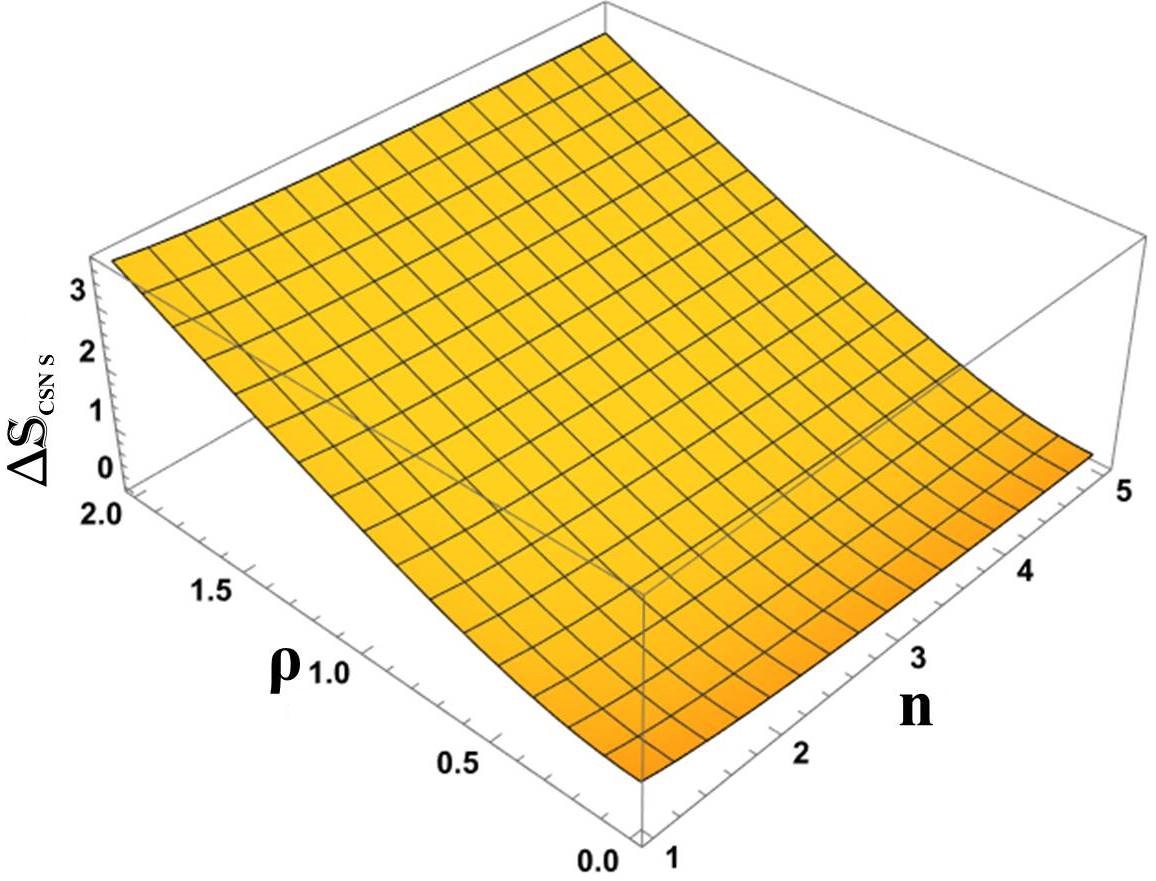}
\caption{3D Plot for $ \Delta {\mathbb{S}}_{CSNS} $ with number state parameter ($ n $) and squeezing parameter ($ \rho $).}
\label{fig:figure_9}
\end{center}
\end{figure}

\begin{figure}[htb]
\begin{center} 
\includegraphics[width=0.68\textwidth]{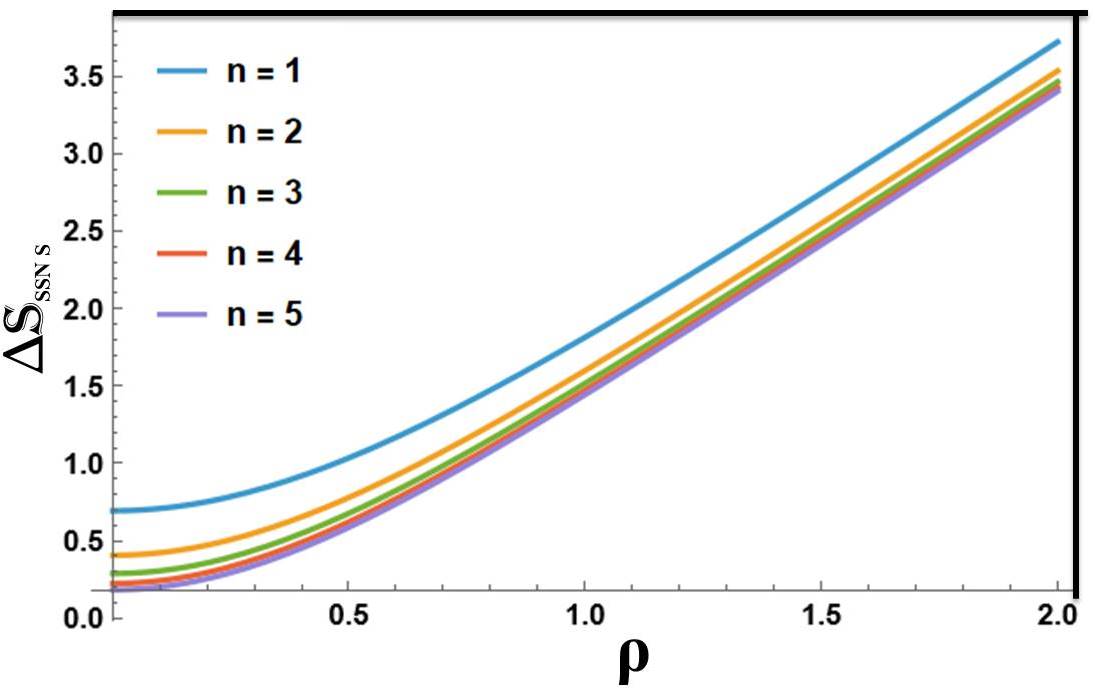}
\caption{2D Plot for $ \Delta {\mathbb{S}}_{CSNS} $ and squeezing parameter ($ \rho $) for different values of number state parameter (n).}
\label{fig:figure_10}
\end{center}
\end{figure}

similarly, variation in mass $\Delta {\mathbb{M}}_{CSNS}$ of the black hole for for single mode  CSNS, using Eqs. (\ref{5.14}, \ref{6.22}) in the Eq. (\ref{6.18})

\begin{equation}\label{6.23}
\Delta {\mathbb{M}}_{CSNS} = {{\omega\over \mathbb{G}^2 \ln \left(1+{1\over\ ({\varUpsilon ^2}+(n)\text{Cosh}(2\rho)+\text{Sinh}^2\rho)}\right)}}~\ln \left[{\varUpsilon ^2}+{(n)\text{Cosh}(2\rho)+\text{Sinh}^2\rho)}\over n \right].
\end{equation}

\begin{figure}[htb]
\begin{center} 
\includegraphics[width=0.68\textwidth]{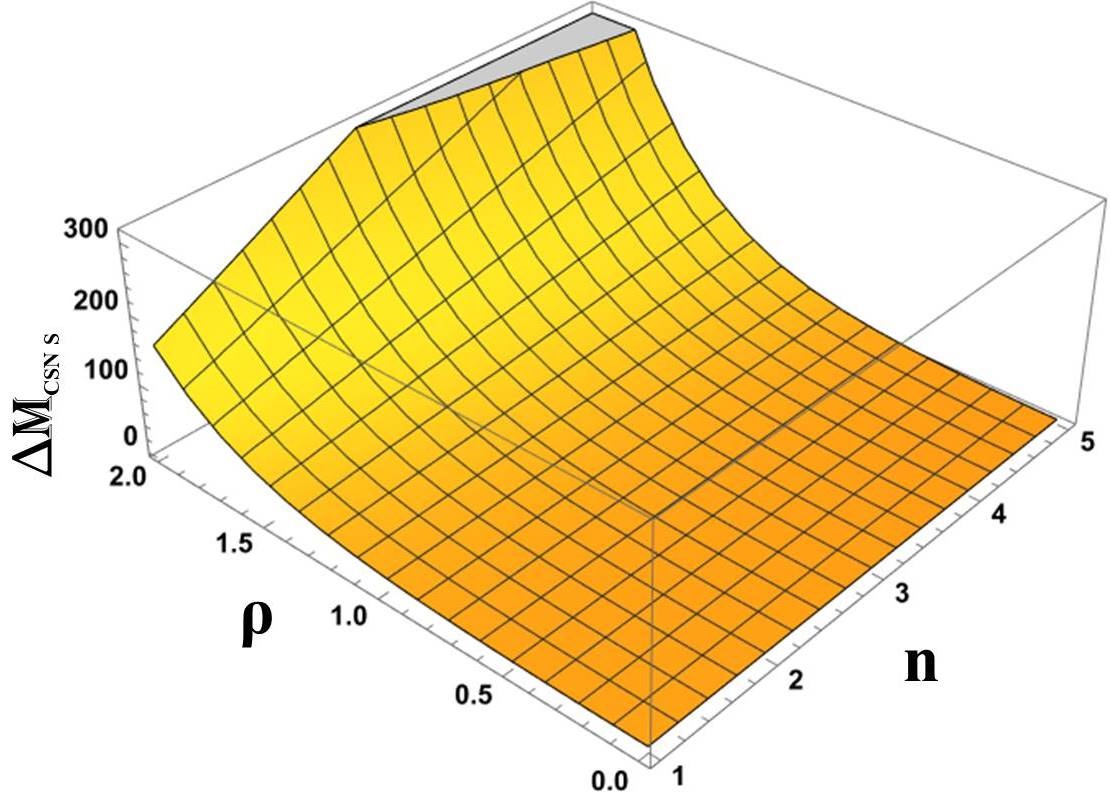}
\caption{3D Plot for $ \Delta {\mathbb{M}}_{CSNS} $ with number state parameter ($ n $) and squeezing parameter ($ \rho $).}
\label{fig:figure_11}
\end{center}
\end{figure}

\begin{figure}[htb]
\begin{center} 
\includegraphics[width=0.68\textwidth]{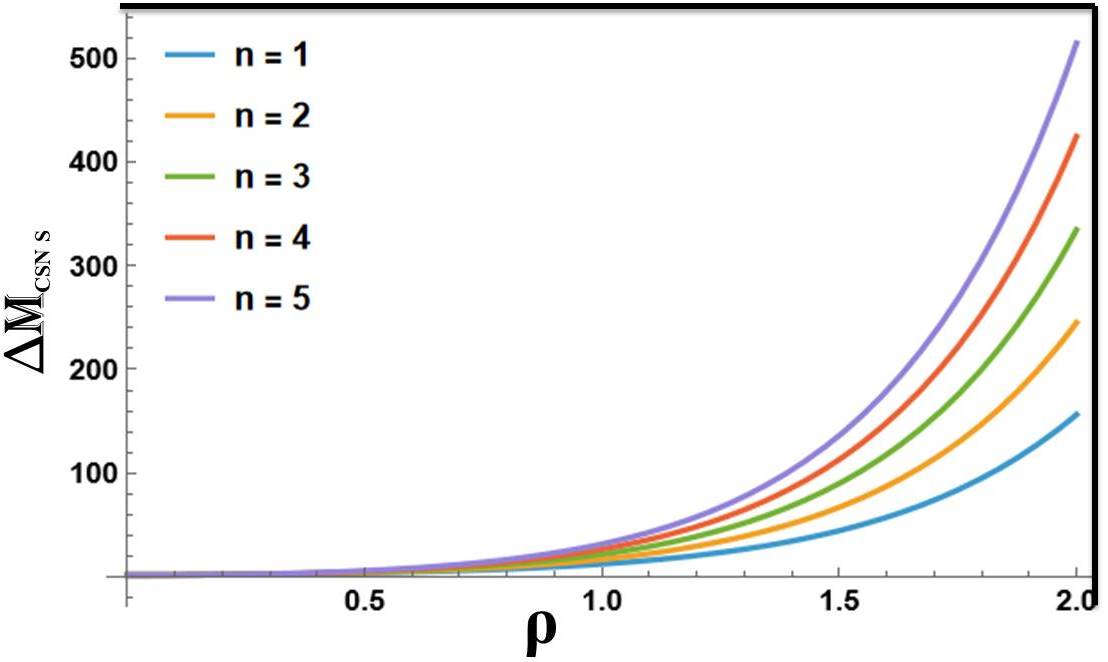}
\caption{2D Plot for $ \Delta {\mathbb{M}}_{CSNS} $ and squeezing parameter ($ \rho $) for different values of number state parameter (n).}
\label{fig:figure_12}
\end{center}
\end{figure}

Now as an important result of the study when $\rho=0$ in Eqs (\ref{6.17} and \ref{6.19}) $\Delta {\mathbb{S}}_{SNS} =0=\Delta {\mathbb{M}}_{SNS}$ and for $\rho=\varUpsilon=0$ in Eqs (\ref{6.22} and \ref{6.23}) the $\Delta {\mathbb{S}}_{CSNS} =0=\Delta {\mathbb{M}}_{CSNS}$ shows there will no change in entropy and mass of black hole i.e.

\begin{equation}\label{6.24}
\Delta {\mathbb{S}}_{SNS}=\Delta {\mathbb{S}}_{CSNS} = \Delta {\mathbb{M}}_{SNS}=\Delta {\mathbb{M}}_{CSNS}=0,
\end{equation}

\section{Discussion and Conclusions }
In this section, we present analysis of Hawking temperature, entropy variation, and black hole mass change for nonclassical inflaton states in the form of SNS and CSNS. The results are based on the analytical expressions derived in earlier sections and are illustrated through twelve figures that systematically explore the dependence of these thermodynamic quantities on the squeezing parameter $\rho$ and number state parameter $n$. Unlike the thermal squeezed state analyses in the literature, the present study focuses exclusively on non thermal, number state resolved quantum states, allowing us to probe distinctly nonclassical corrections to gravitational particle creation.

Figures (\ref{fig:figure_1}-\ref{fig:figure_4}) display the behaviour of the single mode Hawking temperature for SNS (${\mathbb{T}}_{{\mathbb{H}}_{SNS}}$) and CSNS (${\mathbb{T}}_{{\mathbb{H}}_{CSNS}}$). For SNS, figures (\ref{fig:figure_1}-\ref{fig:figure_2}), the temperature exhibits a monotonic enhancement with increasing squeezing parameter $\rho$ for all number state values. The number state dependence is particularly significant: higher values of $n$ shift the temperature upward, demonstrating that number state excitations strongly contribute to the effective temperature of Hawking radiation. This behaviour originates from the nonclassical amplification of field fluctuations, where squeezing and number state occupation jointly increase the expectation values of field variances that feed into the Hawking flux.

For CSNS, figures (\ref{fig:figure_3}-\ref{fig:figure_4}) , similar trends are observed; however, the displacement parameter introduces an additional asymmetry, resulting in a broader rise of temperature for moderate values of $\rho$. This indicates that coherent contributions enhance the particle creation rate even in the presence of strong squeezing. At large squeezing, ${\mathbb{T}}_{{\mathbb{H}}_{CSNS}}$ grows more rapidly than in SNS, highlighting the combined influence of phase sensitive correlations and number state structure in shaping the Hawking spectrum. Overall, the Hawking temperature in both nonclassical states is governed entirely by $\rho$ and $n$, demonstrating that non thermal squeezing provides a robust mechanism for modifying Hawking emission without relying on thermal statistical assumptions.

The entropy variations $\Delta {\mathbb{S}}_{SNS}$ and $\Delta {\mathbb{S}}_{CSNS}$ are shown in figures (\ref{fig:figure_5}-\ref{fig:figure_6}) and figures (\ref{fig:figure_9}-\ref{fig:figure_10}), respectively. For SNS, $\Delta {\mathbb{S}}_{SNS}$ increases with both $\rho$ and $n$, reflecting the growth of information content associated with particle creation. At small squeezing, the entropy exhibits weak variation, but beyond a moderate threshold of $\rho$, the growth becomes nonlinear. This nonlinear rise is indicative of the strong number state dependent squeezing inherent to SNS. Physically, this means that as squeezing strengthens the effective quantum excitation of the field, the entropy produced at the horizon grows substantially, signalling enhanced entanglement between interior and exterior modes.

For CSNS, the behaviour is qualitatively similar but quantitatively more pronounced. Figures (\ref{fig:figure_9}-\ref{fig:figure_10}) show that $\Delta {\mathbb{S}}_{CSNS}$ grows faster with $\rho$ and displays a wider separation between curves for different $n$ compared to SNS. The coherent component introduces additional phase space distortion, generating a higher degree of nonclassicality and consequently larger entropy change. These results reinforce the conclusion that nonclassical correlations, especially those arising from coherent displacements combined with squeezing, significantly amplify entropy generation. This behaviour stands in sharp contrast to thermal squeezed state models, where entropy is predominantly controlled by thermal parameters; in the present non thermal formulation, it is purely the squeezing and number state structure that governs entropy.

Figures (\ref{fig:figure_7}-\ref{fig:figure_8}) and figures (\ref{fig:figure_11}-\ref{fig:figure_12}) illustrate the change in mass $\Delta {\mathbb{M}}$, of black hole due to Hawking emission in SNS and CSNS. Since Hawking radiation leads to mass loss, the magnitude of $\Delta {\mathbb{M}}$ increases with both $\rho$ and $n$. For SNS, figures (\ref{fig:figure_7}-\ref{fig:figure_8}), black hole mass decreases more rapidly for larger $n$, reflecting that number state amplified squeezing accelerates particle emission. As in entropy, a nonlinear dependence on $\rho$ emerges, with mass loss becoming substantial at higher squeezing values.

In the CSNS case figures (\ref{fig:figure_11}-\ref{fig:figure_12}), the mass variation is even more pronounced. The coherent contribution intensifies the effective particle creation rate, producing larger $\Delta {\mathbb{M}}$ across all squeezing scales. For sufficiently large $\rho$, the difference between neighbouring number states becomes significant, confirming that CSNS represents the strongest nonclassical modification of Hawking evaporation among the states considered.

The combined analyses of Hawking temperature, entropy, and mass variation demonstrate that SNS and CSNS introduce significant non thermal modifications to black hole radiation. All thermodynamic quantities depend exclusively on the nonclassical parameters $\rho$ and $n$, confirming that no thermal input is required for particle creation in these states. Compared to earlier thermal squeezed state approaches, the present study reveals a richer structure in the Hawking process driven by number state excitations and genuinely quantum correlations. These results provide a novel perspective on gravitational particle creation, emphasizing the role of nonclassical states in shaping black hole thermodynamics in semiclassical gravity.

\bibliographystyle{unsrt}   % or plain, alpha, abbrv
\bibliography{bibliography}

\end{document}